\documentclass{elsart4-1}

\usepackage{amssymb}

\usepackage[english,francais]{babel}

\usepackage{graphicx,epstopdf,amsfonts,amsmath,array,mathrsfs}
\def\be{\begin{equation}}\def\ee{\end{equation}}
\def\ba{\begin{eqnarray}}\def\ea{\end{eqnarray}}
\def\bmu{\begin{multline}}\def\emu{\end{multline}}
\def\cvp{\raise 2pt\hbox{,}}

\def\tr{\mathop{\rm tr}\nolimits}

\def\diag{\mathop{\rm diag}\nolimits}

\def\d{{\rm d}}\def\nn{{\mathcal N}}
\def\contract#1#2{\overset{\,\hskip #1 pt\lower 8pt\hbox{$\ulcorner$}
\!\!\hrulefill\lower 8pt\hbox{$\negthickspace\urcorner$}\hskip 
-#2 pt\!}}

\def\uN{{\rm U}(N)}

\def\pone{{\mathbb P}^{1}}

\def\pq{(|p\rangle,|q\rangle)}
\def\wl{W_{\rm low}}

\def\La{\Lambda}
\def\la{\lambda}

\def\u{{\rm U}(1)}\def\uu{{\rm U}(2)}
\def\uuu{{\rm U}(3)}\def\uuuu{{\rm U}(4)}

\def\plb#1#2#3{{\it Phys.\ Lett.\ }{\bf B #1} (#2) #3}
\def\npb#1#2#3{{\it Nucl.\ Phys.\ }{\bf B #1} (#2) #3}
\def\npps#1#2#3{{\it Nucl.\ Phys.\ Proc.\ Suppl.\ }{\bf #1} (#2) #3}
\def\prl#1#2#3{{\it Phys.\ Rev.\ Lett.\ }{\bf #1} (#2) #3}
\def\jhep#1#2#3{{\it J. High Energy Phys.\ }{\bf #1} (#2) #3}
\def\prd#1#2#3{{\it Phys.\ Rev.\ }{\bf D #1} (#2) #3}

\def\atmp#1#2#3{{\it Adv.\ Theor.\ Math.\ Phys.\ }{\bf #1} (#2) #3}
\def\cmp#1#2#3{{\it Comm.\ Math.\ Phys.\ }{\bf #1} (#2) #3}
\def\pr#1#2#3{{\it Phys.\ Rep.\ }{\bf #1} (#2) #3}
\def\jmp#1#2#3{{\it J.\ Math.\ Phys.\ }{\bf #1} (#2) #3}

\def\rmp#1#2#3{{\it Rev.\ Mod. Phys. }{\bf #1} (#2) #3}
\newtheorem{e-proposition}[theorem]{Proposition}

\newtheorem{e-definition}[theorem]{Definition\rm}

\renewcommand{\theequation}{\arabic{equation}}
\def\og{\leavevmode\raise.3ex\hbox{$\scriptscriptstyle\langle\!\langle$~}}
\def\fg{\leavevmode\raise.3ex\hbox{~$\!\scriptscriptstyle\,\rangle\!\rangle$}}

\begin{document}
\begin{onecolumn}
\begin{frontmatter}

\hfill\vtop{\hbox{NEIP-04-005}\hbox{LPTENS-04/42}\hbox{hep-th/0410169}}
\vskip -1.3cm

\selectlanguage{english}
\title{Super Yang-Mills, Matrix Models\\ and Geometric Transitions}

\vspace{-2.6cm}

\selectlanguage{francais}
\title{Super Yang-Mills, mod\`eles de matrices\\ et transitions
g\'eom\'etriques}


\selectlanguage{english}
\author[1,2]{Frank Ferrari}
\ead{frank.ferrari@ulb.ac.edu}

\address[1]{Institut de Physique, Universit\'e de Neuch\^atel\\
rue A.-L.\ Br\'eguet 1, CH-2000 Neuch\^atel, Suisse}
\address[2]{Service de Physique Th\'eorique et Math\'ematique\\
and International Solvay Institutes\\
Universit\'e Libre de Bruxelles, Campus de la Plaine\\
CP 231, Blvd.\ du Triomphe, B-1050 Bruxelles, Belgique}

\begin{abstract}
I explain two applications of the relationship between four
dimensional $\nn=1$ supersymmetric gauge theories, zero dimensional
gauged matrix models, and geometric transitions in string theory. The
first is related to the spectrum of BPS domain walls or BPS branes. It
is shown that one can smoothly interpolate between a D-brane state,
whose weak coupling tension scales as $N\sim 1/g_{\mathrm s}$, and a
closed string solitonic state, whose weak coupling tension scales as
$N^{2}\sim 1/g_{\mathrm s}^{2}$. This is part of a larger theory of
$\nn=1$ quantum parameter spaces. The second is a new purely geometric
approach to sum exactly over planar diagrams in zero
dimension. It is an example of open/closed string duality.

\vskip 0.5\baselineskip

\selectlanguage{francais}
\noindent{\bf R\'esum\'e}
\vskip 0.5\baselineskip
\noindent
J'expose deux applications de la relation entre les th\'eories de
jauge supersym\'etriques $\nn=1$ \`a quatre dimensions, les mod\`eles
de matrices jaug\'es \`a z\'ero dimension, et les transitions
g\'eom\'etriques en th\'eorie des cordes. La premi\`ere traite du
spectre des parois ou branes BPS. Il est d\'emontr\'e qu'un \'etat de type
D-brane, dont la tension en couplage faible est proportionnelle \`a
$N\sim 1/g_{\mathrm s}$, et un \'etat de type soliton de corde
ferm\'ee, dont la tension en couplage faible est proportionnelle \`a
$N^{2}\sim 1/g_{\mathrm s}^{2}$, peuvent \^etre continument
d\'eform\'es l'un en l'autre. Ceci est un aspect de la th\'eorie plus
g\'en\'erale des espaces quantiques de param\`etres $\nn=1$. La
deuxi\`eme est une nouvelle m\'ethode, purement g\'eom\'etrique, pour
calculer exactement la somme sur les diagrammes planaires en z\'ero
dimension. C'est un exemple de dualit\'e entre cordes ouvertes et cordes
ferm\'ees.


\noindent{\small{\it Mots-cl\'es~:} Th\'eories de jauge
supersym\'etriques~; Mod\`eles de matrices~; Th\'eorie des cordes}
\vskip 0.5\baselineskip\noindent
To appear in the proceedings of the Strings 2004 conference in Paris, 
published in the Comptes Rendus de l'Acad\'emie des Sciences.
\end{abstract}
\end{frontmatter}
\end{onecolumn}
\begin{twocolumn}
\selectlanguage{francais}
\noindent{\bf Version fran\c{c}aise abr\'eg\'ee}
\vskip .6cm

L'\'etude des th\'eories de jauge supersym\'etriques en couplage fort
a connu deux avanc\'ees majeures ces dix derni\`eres ann\'ees. L'une
est fond\'ee sur la notion de dualit\'e S et sur l'utilisation des
propri\'et\'es d'analyticit\'e de certaines observables pour obtenir
des r\'esultats exacts \cite{MO,Sen,SW,N1}. L'autre est fond\'ee sur
la construction de th\'eories de cordes, math\'ematiquement
\'equivalentes aux th\'eories de jauge, mais plus simples \`a
\'etudier dans le r\'egime de couplage fort \cite{mal}. Le point de
contact entre ces deux d\'eveloppements \cite{holtop,Gtr,DV,ferrev}
est \`a l'origine d'une activit\'e intense depuis 2002. Nous
souhaitons pr\'esenter deux aspects compl\'ementaires de ces
recherches, \`a la suite de travaux publi\'es dans \cite{fer1,fer2}.

\vskip .5cm
\noindent{\it Parois}
\vskip .4cm

Le premier sujet concerne le spectre des parois (ou membranes) BPS
dans les th\'eories $\nn=1$. Dans la th\'eorie de jauge pure $\uN$, ces
parois existent en raison de la brisure dynamique de la sym\'etrie
chirale $\mathbb Z_{2N}\rightarrow\mathbb Z_{2}$. Elles ont \'et\'e
\'etudi\'ees, par exemple, par Witten \cite{Witwalls}. Elles se
comportent comme des deux-branes de Dirichlet: leur tension est
proportionnelle \`a $N\sim 1/g_{\mathrm s}$, et des cordes ouvertes
peuvent s'y attacher. Lorsque l'on rajoute \`a la th\'eorie un
multiplet chiral de mati\`ere dans la repr\'esentation adjointe, avec
un superpotentiel $W$ \`a l'ordre des arbres tel que (\ref{dWspec}),
il appara\^\i t un nouveau type de parois qui interpolent entre les
vides $\langle X\rangle = x_{+}$ et $\langle X\rangle = x_{-}$. Ces
parois peuvent \^etre \'etudi\'ees en d\'etail dans la th\'eorie
classique. Elles se comportent comme des deux-branes solitoniques,
avec une tension proportionnelle \`a $N^{2}\sim 1/g_{\mathrm s}^{2}$.

Il est possible de calculer exactement les tensions de ces diverses
parois ou membranes, en utilisant soit le formalisme du mod\`ele de
matrice \cite{DV} en th\'eorie des champs, soit celui de la transition
g\'eom\'etrique en th\'eorie des cordes \cite{Gtr}. Le r\'esultat
\cite{fer1}, de nature purement non-perturbative, est explicit\'e dans
les formules (\ref{wlowex},\ref{T1},\ref{T2}). La propri\'et\'e
fondamentale de ces formules \cite{fer1} est de montrer qu'il est
possible de d\'eformer de mani\`ere continue les membranes de type
Dirichlet en membrane de type soliton et vice-versa (voir Figure 1).
De plus, pour certaines valeurs particuli\`eres des param\`etres, la
tension de la membrane peut \^etre proportionnelle \`a une puissance
fractionnaire de $N$.

Ces propri\'et\'es de continuation analytique sont tr\`es
g\'en\'erales et ont des cons\'equences drastiques sur la structure de
l'espace quantique des param\`etres des th\'eories $\nn=1$
\cite{fer1,fer4,CSW}. En particulier, cet espace, qui a de multiples
composantes disconnect\'ees au niveau classique, s'av\`ere \^etre
connexe dans tous les cas \'etudi\'es lorsque les corrections
quantiques non-perturbatives sont prises en compte \cite{fer1,fer4}.
Le r\'esultat dans le cas de la th\'eorie de jauge $\uuuu$ est
repr\'esent\'e sur la Figure 2.

L'un de nos r\'esultats les plus important a \'et\'e de montrer que
l'on pouvait d\'eformer continument une D-brane en une brane
solitonique et vice-versa. Nous avons aussi d\'ecouvert de nouveaux
objets \'etendus avec une tension proportionnelle \`a une puissance
fractionnaire de la constante de couplage de corde. Un point
int\'eressant est la possibilit\'e de construire des branes
solitoniques dans le cadre d'une th\'eorie de jauge classique. Une
description explicite de la th\'eorie microscopique sur la membrane
peut alors \^etre donn\'ee en principe par la quantification
semiclassique. Une telle description microscopique n'a pas pu \^etre
obtenue jusqu'\`a pr\'esent dans le cadre de la th\'eorie des cordes.

\vskip .6cm
\noindent{\it Diagrammes planaires et g\'eom\'etrie}
\vskip .45cm

Nous avons deux techniques \`a notre disposition pour calculer le
potentiel effectif d'une th\'eorie de jauge $\nn=1$. La premi\`ere
consiste \`a utiliser le mod\`ele de matrice associ\'e (\ref{MM}),
l'autre consiste \`a construire la th\'eorie de jauge sur une D-brane
de la th\'eorie des cordes puis \`a utiliser la dualit\'e entre cordes
ouvertes et cordes ferm\'ees. L'\'equivalence des deux proc\'edures
implique que les int\'egrales (\ref{MM}) peuvent \^etre calcul\'ees
dans la limite planaire par une m\'ethode purement
g\'eom\'etrique, que nous nous proposons de d\'ecrire \cite{fer2}.

Le point de d\'epart est la vari\'et\'e de Calabi-Yau $\mathscr M$,
couverte par deux cartes et d\'efinie par les fonctions de transition
(\ref{tf}). On peut montrer que cette vari\'et\'e non-compacte
contient des deux-sph\`eres holomorphes autour desquelles peuvent
\^etre enroul\'ees des D5-branes de la th\'eorie de type IIB. La
th\'eorie \`a quatre dimensions qui en r\'esulte est $\nn=1$ avec
groupe de jauge $\uN$, $M$ multiplets chiraux dans la repr\'esentation
adjointe, et un superpotentiel \`a l'ordre des arbres donn\'e par
(\ref{Wfor}) \cite{fer2}. Ce superpotentiel est le plus g\'en\'eral
jamais construit pour des th\'eories de ce type. Par exemple, pour
$M=2$, on peut obtenir un superpotentiel arbitraire $W(x,y)$ dans le
cas $N=1$ o\`u les deux matrices $X=X_{1}$ et $Y=X_{2}$ commutent.
Dans le cas g\'en\'eral, la formule (\ref{Wfor}) indique que l'ordre
des termes \`a adopter correspond \`a celui introduit par Weyl en
m\'ecanique quantique (\ref{Weyl}). La m\'ethode g\'eom\'etrique que
nous \'etudions doit permettre, en principe, de r\'esoudre tous les
mod\`eles du type (\ref{Wfor}), un r\'esultat qui semble impossible
\`a obtenir par les techniques traditionnelles.

En raison de la relation (\ref{gYM}) entre couplage de Yang-Mills et
volume des deux-sph\`eres, on peut s'attendre \`a ce que les
deux-sph\`eres sur lesquelles s'enroulent les D5-branes aient tendance
\`a r\'etr\'ecir, puis \`a dispara\^\i tre, lorsque l'effet du groupe de
renormalisation est pris en compte. Math\'ematiquement, cette
transformation est d\'ecrite par une application birationnelle $\pi$
(\ref{bdm}) qui transforme l'espace $\mathscr M$ en un espace
singulier ${\mathscr M}_{0}$ (l'application inverse est en g\'en\'eral
appel\'ee \'eclatement). Le nouvel espace de Calabi-Yau ${\mathscr
M}_{0}$ est en tout point semblable \`a l'espace de d\'epart $\mathscr
M$, si ce n'est que les deux-sph\`eres de $\mathscr M$ sont
remplac\'ees par des points singuliers de ${\mathscr M}_{0}$. Le
calcul de $\pi$, duquel se d\'eduit imm\'ediatement celui de
${\mathscr M}_{0}$, n'est pas syst\'ematique et peut pr\'esenter des
difficult\'es. Un point remarquable est que la correspondance avec le
mod\`ele de matrice implique que la vari\'et\'e ${\mathscr M}_{0}$
code compl\`etement la th\'eorie des repr\'esentations des \'equations
matricielles (\ref{cleq}). Ceci est un r\'esultat puissant
et inattendu pour les repr\'esentations de dimensions sup\'erieures ou
\'egales \`a deux. Il montre que la g\'eom\'etrie classique est
capable de rendre compte de la structure non-commutative qui
appara\^\i t lorsque l'on consid\`ere des D-branes. Cette
propri\'et\'e peut \^etre v\'erifi\'ee sur des exemples particuliers
au prix de calculs d\'etaill\'es \cite{fer2}.

Pour r\'esoudre le mod\`ele de matrice (\ref{MM}) quand $S\not =0$,
une \'etape suppl\'ementaire est n\'ecessaire au cours de laquelle
l'espace ${\mathscr M}_{0}$ est d\'eform\'e en un nouveau Calabi-Yau
non-singulier $\hat{\mathscr M}$. Les d\'eformations autoris\'ees sont
soumises \`a certaines contraintes de normalisabilit\'e d\'ecrites
dans \cite{fer2}. Le nombre $\mathscr N_{\mathrm{geo}}$ de
param\`etres ind\'ependants qui en r\'esultent doit \^etre \'egal au
nombre de param\`etres $\mathscr N_{\mathrm{alg}}$ dans le mod\`ele de
matrices. Ces derniers sont associ\'es au choix d'une solution
classique \`a partir de laquelle les diagrammes de Feynman planaires
sont d\'efinis, et leur nombre est donc \'egal au nombre de
repr\'esentations irr\'eductibles des \'equations (\ref{cleq}).
L'identit\'e (\ref{geomalg}) traduit cette relation profonde entre
g\'eom\'etrie classique et alg\`ebre non-commutative. Elle est
satisfaite par tous les exemples \'etudi\'es \cite{fer2}. \`A partir
de la connaissance de $\hat{\mathscr M}$, on peut aussi calculer
explicitement les r\'esolvantes (\ref{gdef}) pour les diff\'erentes
matrices du mod\`ele en utilisant (\ref{disc}), ainsi que la fonction
de partition \cite{fer2}.

De nombreuses questions n'ont pas encore \'et\'e \'elucid\'ees dans
cette correspondance entre mod\`eles de matrices et g\'eom\'etrie. Par
exemple, on peut se demander quelles sont les propri\'et\'es
particuli\`eres des mod\`eles (\ref{Wfor}) qui font qu'ils puissent
\^etre reli\'es \`a une g\'eom\'etrie de Calabi-Yau (th\'eorie des
repr\'esentations de (\ref{cleq}), \'equations de boucle, etc...). Le
probl\`eme ouvert le plus important est de comprendre si l'application
$\pi$ existe toujours pour la g\'eom\'etrie (\ref{tf}). Fournir une
preuve d'existence rigoureuse est un probl\`eme math\'ematique
difficile \cite{kol}. Si l'on pouvait trouver un algorithme permettant
de construire $\pi$, on pourrait alors r\'esoudre automatiquement tous
les mod\`eles de matrices (\ref{Wfor}). Si, par contre, l'application
$\pi$ n'existait pas en g\'en\'eral, alors le cadre des transitions
g\'eom\'etriques serait insuffisant pour d\'ecrire nos mod\`eles. Ceci
aurait des cons\'equences fondamentales pour notre compr\'ehension des
vides $\nn=1$ en th\'eorie des cordes.

\selectlanguage{english}

%
\section{Introduction}

In the last decade, a wealth of exact results in supersymmetric gauge
theories have been derived. Beyond the original physical insight based
on S-duality \cite{MO,Sen,SW}, a great variety of ideas and techniques
have been used: singularity theory \cite{ST}, integrable systems
\cite{IS}, geometric engineering \cite{GE}, brane constructions
\cite{BC}, etc\dots The common thread of these developments is that
they deal with observables that depend holomorphically on the
couplings/moduli/parameters of the problem. This includes the low
energy effective actions of $\nn=2$ theories \cite{SW,SW2,N2} and the
effective superpotentials and $\u$ couplings of $\nn=1$ theories
\cite{N1}. The set of stable BPS states in $\nn=2$ theories can also
be determined in this framework \cite{BPS}.

Another major breakthrough has been the explicit construction
of gauge theory/string theory dual pairs \cite{mal}. This open/closed
string duality is in principle exact. However, most of the useful
calculations can only be performed approximately, and are valid in a
regime where either the closed string theory is tractable (small
string coupling or large $N$, and small curvature or large $g_{\mathrm
YM}^{2}N$), or the open string theory is tractable (small $g_{\mathrm
YM}^{2}N$, which is the usual Feynman perturbation theory).

The point of contact between the two above-mentioned line of
developments has generated an intense activity in the last couple of
years \cite{ferrev}. The basic observation is that the holomorphic
observables in gauge theories correspond to the topological sector of
the string theories \cite{holtop}. The open/closed string duality can
be implemented exactly in the context of the topological strings,
using the idea of geometric transitions \cite{Gtr}. It turns out that
the result is encoded in the sum over planar diagrams of a particular
zero-dimensional gauged matrix model \cite{DV}. This provides a
beautifully simple and unified framework to rederive many of the
already known exact results. It also opens up new directions of
research. The aim of this talk is to discuss two complementary aspects
of this subject, based on work first published in \cite{fer1,fer2}.

We shall consider four dimensional $\nn=1$ super Yang-Mills theories
with gauge group $\uN$ and $M$ matter chiral multiplets $X_{i}$ in the
adjoint representation. Asymptotic freedom is guaranteed for $M\leq
2$. The case $M=3$ includes the conformally invariant $\nn=4$ theory.
The cases $M>3$ can be defined by embedding the gauge theory
in string theory, which provides a natural UV cut-off $\La_{0}$.
We consider tree-level superpotentials for
the matter fields of the form
\be\label{superpot}W = \tr V(X_{i})\, ,\ee
where $V$ is a polynomial. Renormalizability is not an issue, because
the holomorphic datas ($F$-terms) cannot depend on the UV cut-off
$\La_{0}$. Equivalently, we can always introduce auxiliary fields to
make $W$ at most cubic. Our super Yang-Mills theories can be
geometrically engineered by considering a stack of $N$ D5 branes in
type IIB string theory on $\mathbb R^{4}\times\mathscr M$. The space
$\mathscr M$ is a non-compact Calabi-Yau threefold. The D5 branes
span $\mathbb R^{4}$ and wrap two-spheres $\mathbb
P^{1}\subset\mathscr M$. We shall explain in Section 3 how to choose
the space $\mathscr M$ to get an arbitrary number $M$ of adjoint
chiral superfields and a large class of superpotentials $W$ of the
form (\ref{superpot}).

From the gauge theory point of view, the quantum corrections to
$F$-terms are computed, according to Dijkgraaf and Vafa \cite{DV}, by
summing up the planar diagrams of a zero dimensional gauged matrix
model
\be\label{MM}
\int_{\mathrm{planar}}\!\! e^{-\frac{n}{S}W}\,\prod_{i}\d X_{i}\, .\ee
The coupling $S$ of the matrix model is identified with the glueball
superfield of the gauge theory,
\be\label{Sdef}S=-\frac{\tr
W^{\alpha}W_{\alpha}}{16\pi^{2}N}\,\cdotp\ee
From the string theory point of view, the quantum corrections 
are obtained by using the open/closed string duality. Following
\cite{Gtr}, the closed string dual is type IIB on a new non-compact
Calabi-Yau $\hat{\mathscr M}$, with no D-branes, but with three-form flux 
turned on (consistently with the original D-branes charges).
The space $\hat{\mathscr M}$ is a deformation of the singular
Calabi-Yau obtained by blowing down $\mathscr M$. Examples are
presented in Section 3.

We thus have two powerful tools to compute the exact effective
superpotentials of the gauge theories. An interesting consequence of
the results so obtained is described in Section 2. In Section 3 we use
the fact that the two approaches, gauge-theoretic and
string-theoretic, must be equivalent, to describe a new purely
geometric method to sum up planar diagrams in zero dimension.

\section{Domain walls}
\subsection{Pure $\uN$ super Yang-Mills}

The pure $\uN$ super Yang-Mills theory has a $\mathbb Z_{2N}$ chiral
symmetry that acts on the gluino field, $\lambda_{\alpha}\rightarrow
e^{i\pi k/N}\lambda_{\alpha}$. The glueball superfield (\ref{Sdef}) is
a good, gauge invariant, order parameter for this symmetry. It turns out
that strong quantum effects break $\mathbb Z_{2N}$ down to $\mathbb
Z_{2}$. This breaking implies an $N$-fold degeneracy of the space of
vacua, which are denoted by $|p\rangle$, $p\in\mathbb Z_{N}$. In the 
$p^{\mathrm{th}}$ vacuum,
\be\label{gvev}\langle p|S|p\rangle = \frac{1}{16\pi^{2}N}\langle p|
\tr\lambda^{\alpha}\lambda_{\alpha}|p\rangle=\La^{3}e^{2i\pi k/N}\, ,\ee
where $\La$ is a conveniently normalized dynamically generated complex
mass scale. The existence of a discrete set of degenerate vacua
implies the existence of domain walls $(|p\rangle,|q\rangle)$
interpolating between them. These domain walls are discussed for
example in \cite{Witwalls}. They are BPS two-branes, which implies in
particular that their tension can be expressed in terms of the quantum
effective superpotential,
\be\label{Tform} T_{\pq} =
N\bigl|\wl^{|p\rangle}-\wl^{|q\rangle}\bigr|\, ,\ee
with
\be\label{wlpure}\wl^{|p\rangle} = N\La^{3}e^{2i\pi p/N}\, .\ee
For ``elementary'' domain walls, that join the $p^{\mathrm{th}}$ and
the $(p+1)^{\mathrm{th}}$ vacua, (\ref{Tform}) yields
\be\label{Telem}T_{(|p\rangle,|p+1\rangle)} =
2N^{2}\La^{3}\sin\frac{\pi}{N}\,\cdotp\ee
In the large $N$ or small string coupling limit, the tension scales as
$N\sim 1/g_{\mathrm s}$. This is the behaviour expected for a D-brane,
and it was indeed argued in \cite{Witwalls} that open strings can end
on these domain walls. It is in principle very difficult to study
these walls in the framework of gauge theory, because they belong
entirely to the strong coupling regime of the theory. On the other
hand, we see that in principle a simple description in terms of open
strings exists. Of course, the explicit construction of these open
strings remains an outstanding unsolved problem.

\subsection{$\uN$ super Yang-Mills with one adjoint $X$}

Let us now add one adjoint chiral multiplet $X$ to the pure theory,
with a tree level superpotential $W$ such that 
\be\label{dWspec}\d W = g\tr\bigl[(X-x_{+})(X-x_{-})\,\d X\bigr]\, .\ee
We now have $\frac{1}{6} N(N^{2}+11)$ vacua corresponding to various
patterns of gauge and chiral symmetry breaking. For example, for
$N=2$, we have two chirally asymmetric vacua similar to the pure gauge
theory case and corresponding to $\langle X\rangle_{\mathrm{cl}} =
x_{+} I_{2}$, two other similar vacua corresponding to $\langle X
\rangle_{\mathrm{cl}} = x_{-} I_{2}$, and a Coulomb phase vacuum
corresponding to $\langle X\rangle_{\mathrm{cl}} =\diag(x_{+},x_{-})$
and pattern of gauge symmetry breaking $\uu\rightarrow\u\times\u$.

We focus on the $2N$ vacua $|p,\pm\rangle$ corresponding to $\langle
X\rangle_{\mathrm{cl}} = x_{\pm} I_{2}$ and for which the low energy
theory is the pure $\uN$ theory with scale $\La$. We now have to
consider two types of domain walls. The first type corresponds to
walls interpolating between the vacua $|p,+\rangle$ and $|q,+\rangle$
or $|p,-\rangle$ and $|q,-\rangle$. These are the D-brane states of
the pure low energy Yang-Mills theory discussed before. The second
type corresponds to walls interpolating between the vacua
$|p,-\rangle$ and $|q,+\rangle$. These walls exist in the classical
theory, and in particular they can be quantized semiclassically. It is
straightforward to show that they are BPS and to compute their tension
in the classical limit. The result shows that the tension scales as
$N^{2}\sim 1/g_{\mathrm s}^{2}$. These domain walls are thus closed
string solitons.

The spectrum of extended objects in the gauge theory with one adjoint
is thus very similar to the spectrum in ordinary supersymmetric string
theories: we have both D-branes and closed string solitons.

\subsection{Interpolation}

We now come to our main point. For concreteness, let us specialize to
\be\label{Wspec2} W = \tr\Bigl( \frac{1}{2} mX^{2} + \frac{1}{3} g
X^{3}\Bigr)\, ,\ee
for which
\be\label{xpmspec} x_{+} = 0\, ,\quad x_{-}=-m/g\, .\ee
If the dynamically generated scale of the full theory is $\La_{\mathrm
u}$, then the low energy scales $\La_{\pm}$ in the vacua
$|p,\pm\rangle$ are given by
\be\label{lesca}\La_{\pm}^{3} = \pm m\La_{\mathrm u}^{2}=\pm\La^{3}\, .\ee
Let us introduce the dimensionless parameter
\be\label{ldef} \la = \frac{8g^{2}\La^{3}}{m^{3}}\,\cdotp\ee
The effective superpotentials can be computed exactly \cite{fer3} in
the various vacua $|p,\pm\rangle$, by using for example the
Dijkgraaf-Vafa matrix model prescription. The result is
\be\label{wlowex}\wl^{|p,\pm\rangle} =
\frac{2N}{3}\La^{3}\,\frac{1}{\la}\Bigl[ 1\mp\bigl( 1-\la e^{2i\pi
p/N}\bigr)^{3/2}\Bigr]\, .\ee
This is a highly non-perturbative formula. The weak coupling
expansions, in powers of $\la\propto\La^{3}$, are fractional instanton 
series. The leading terms yield
\be\label{wcexp}\wl^{|p,\pm\rangle} = W_{\mathrm{low,\,
cl}}^{|\pm\rangle} +
N\La_{\pm}^{3}e^{2i\pi p/N} + \mathcal{O}(\La^{6}/m^{3})\, ,\ee
consistently with (\ref{wlpure}). In particular, the tension of the
solitonic branes in the classical limit is given by
\be\label{Tclsol}T_{\mathrm{cl}} = N\bigl| W_{\mathrm{low,\,
cl}}^{|+\rangle} - W_{\mathrm{low,\, cl}}^{|-\rangle}\bigr| =
N^{2}\,\Bigl|\frac{m^{3}}{6g^{2}}\Bigr|\, ,\ee
which is proportional to $N^{2}$ as already advertised.
More generally, the exact tensions of the various branes are given in 
terms of (\ref{wlowex}) by
\begin{align}\label{T1}T_{(|p,\pm\rangle,|q,\pm\rangle)} &=
N\bigl| \wl^{|p,\pm\rangle} - \wl^{|q,\pm\rangle}\bigr|\\ \label{T2}
T_{(|p,\pm\rangle,|q,\mp\rangle)} &=
N\bigl| \wl^{|p,\pm\rangle} - \wl^{|q,\mp\rangle}\bigr|\, .\end{align}

A very important and general feature of the fractional instanton
series for $\wl$ is that they have a finite radius of convergence.
This implies that analyticity is lost at special points and that $\wl$
is a multivalued function of the parameters. In our example
(\ref{wlowex}), $\wl^{|p,\pm\rangle}$ is a two-sheeted function.
Analyticity is lost at the square root branching point
\be\label{bp} \la = \la_{p} = e^{-2i\pi p/N}\, .\ee
By going through the branch cut, we flip the sign of the square root,
and thus permute $\wl^{|p,+\rangle}$ and $\wl^{|p,-\rangle}$. For
example, by following the path in parameter space depicted in Figure
1, we join smoothly a weakly coupled region in the vacuum
$|p,+\rangle$ to a weakly coupled region in the vacuum $|p,-\rangle$.
Exchanging a $|+\rangle$ vacuum with a $|-\rangle$ vacuum by smoothly
varying the parameters is not an exploit. This can be achieved
classically by permuting the roots $x_{+}$ and $x_{-}$. The highly
non-trivial feature is that we are able to exchange $|p,+\rangle$ and
$|p,-\rangle$, {\it without changing any of the other vacua
$|q,\pm\rangle$ for $q\not = p$} \cite{fer1}. This is a
non-perturbative property, made possible by the fact that the
branching points (\ref{bp}) and associated branch cuts for the vacua
with different values of $p$ are at different locations.

\begin{figure}
\centerline{\includegraphics[width=4.5truecm]{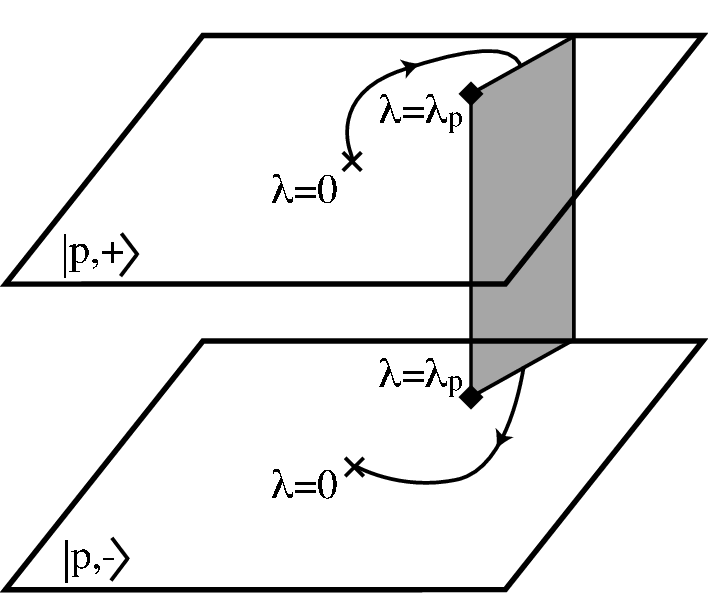}}
\caption{A path in parameter space interpolating smooothly between the
vacua $|p,+\rangle$ and $|p,-\rangle$. The other vacua $|q,\pm\rangle$
for $q\not = p$ are not changed.
\label{fig1}}
\end{figure}

The physics underlying the smooth interpolation 
\be\label{sm1} |p,+\rangle\rightarrow |p,-\rangle\ee
depicted in Figure 1 is best illustrated by looking at the domain
walls. Consistently with the multivaluedness of the tension formulas
(\ref{T1},\ref{T2},\ref{wlowex}), a D-brane is smoothly deformed into
a solitonic brane,
\be\label{sm2} (|p,+\rangle,|p+1,+\rangle)\rightarrow
(|p,-\rangle,|p+1,+\rangle)\, ,\ee
and vice versa. In other words, the multivaluedness implies that there
are several possible classical limits, some for which the tension
scales as $N$, and others for which the tension scales as $N^{2}$.
Obviously, at strong coupling, the distinction between D-branes and
solitons does not make sense. This is particularly striking at the
point $\la=\la_{p}$, for which
\begin{align}\label{Tcrit}T_{(|p,+\rangle,|p+1,+\rangle)}(\la_{p})&= 
T_{(|p,-\rangle,|p+1,+\rangle)}(\la_{p})\\ \label{sqrtN}
&\hskip -.2cm\mathop{\propto}_{N\rightarrow\infty}\sqrt{N}\sim
1/\sqrt{g_{\mathrm s}}\, .\end{align}
We have discovered a new type of extended object in the theory at
$\la=\la_{p}$, with a tension proportional to
$\sqrt{N}\sim\sqrt{1/g_{\mathrm s}}$ \cite{fer1}. Other fractional
powers can be achieved by looking at more complicated branching points
\cite{fer1}.

\subsection{Quantum parameter space}
\begin{figure*}
\centerline{\includegraphics[width=14.15truecm]{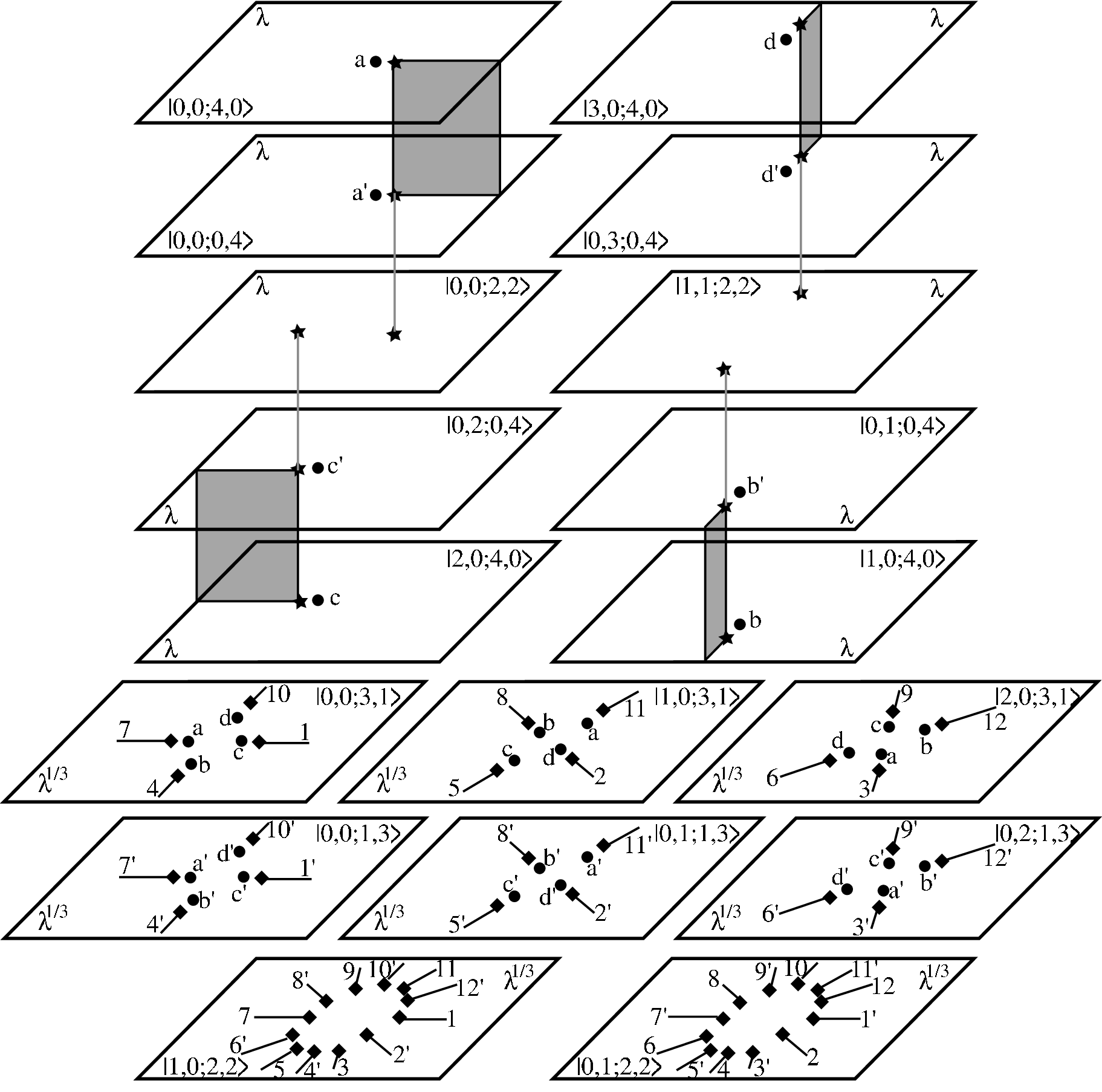}}
\caption{Sketch of the quantum parameter space $\mathcal M_{\mathrm
q}$ for gauge group $\uuuu$. The sheets are parametrized by $\la$ or
$\la^{1/3}$. The vacua for which $X$ has $N_{1}$ eigenvalues at zero
and $N_{2}$ eigenvalues at $-m/g$ are denoted by
$|k_{1},k_{2};N_{1},N_{2}\rangle$, with $k_{j}\in\mathbb Z_{N_{j}}$.
The dots, squares and stars represent massless monopole points (at
$\la = (4/5)e^{-ik\pi /2}$), branching points (at
$\la^{1/3}=(3^{1/4}/2^{1/6})e^{-ik\pi/6}$), or both (at $\la =
e^{-ik\pi/2}$), respectively. Due to the complexity of the diagram, we
have not been able to represent explicitly all the identifications
between sheets. It is understood that singularities and branch cuts
with the same label are identified.
\label{fig3}}
\end{figure*}

The physics discussed above is a basic consequence of a very general
structure \cite{fer1,fer4,CSW}. Let us consider the space, that we
call the classical parameter space $\mathcal M_{\mathrm{cl}}$, whose
points are labeled by the parameters of the gauge theory {\it and} by
a given vacuum. The space $\mathcal M_{\mathrm{cl}}$ has many
disconnected components, or sheets, each labeled by a
vacuum. For example, in the theory (\ref{dWspec}), $\mathcal
M_{\mathrm{cl}}$ has $\frac{1}{6}N(N^{2}+11)$ components parametrized
by $\lambda$. In the full quantum theory, $\wl$ is multivalued and the
associated branch cuts glue some of the sheets together. The resulting
space is called the quantum parameter space $\mathcal M_{\mathrm q}$
\cite{fer1,fer4}. The effective superpotential $\wl$ is always a
single-valued function on $\mathcal M_{\mathrm q}$.

For example, in the $\uu$ theory, $\mathcal M_{\mathrm{cl}}$ has five
disconnected components corresponding to the four confining vacua
$|0,+\rangle$, $|1,+\rangle$, $|0,-\rangle$, $|1,-\rangle$ and the
Coulomb vacuum $|\mathrm{C}\rangle$. From the above discussion, we
know that the sheets labeled by $|0,+\rangle$ and $|0,-\rangle$ on the
one hand, and $|1,+\rangle$ and $|1,-\rangle$ on the other hand, are
glued together. The Coulomb sheet cannot be smoothly connected to the
other sheets, because it is in a different phase. However, there is a
connection at points where monopoles become massless. The phase
transition is triggered by the condensation of the massless monopoles
through a standard Higgs mechanism in magnetic variables. Taking into
account both the smooth interpolations through branch cuts and the
phase transitions through massless monopole points, we get a fully
connected quantum parameter space $\mathcal M_{\mathrm q}$
\cite{fer1}.

The $\nn=1$ quantum parameter spaces are reminiscent of the $\nn=2$
quantum moduli spaces, but there are important differences. The most
notable is that motion on parameter space is not associated with
a massless scalar, which makes the $\nn=1$ models much more
realistic. Moreover, the quantum corrections in $\nn=1$ have even more
drastic effects than in $\nn=2$, since they change the topology of the 
classical space in addition to producing monopole singularities.

In all cases that have been studied so far, $\mathcal M_{\mathrm q}$
turns out to be fully connected. It is not known whether this is a
general property. For example, the rather intricate case of $\uuuu$,
with classically eighteen disconnected components, was worked out in
\cite{fer4}. The result is depicted in Figure 2. There are ten sheets
in a confining phase and eight sheets in a Coulomb phase. The unbroken
gauge group in the Coulomb phase can be either $\uu\times\uu$ (two
vacua) or $\uuu\times\u$ (six vacua). To show that these eight sheets
can be smoothly connected, we can study the analytic structure of
$\wl$ as already explained. It is often more convenient to work with
the derivatives of $\wl$, that yield the expectation values of various
chiral operators. Remarquably, these vevs are the solutions of
algebraic equations. In our case, it is natural to consider
\be\label{defu} u = -\frac{g}{m}\langle\tr X\rangle\, .\ee
The possible classical values of $u$ in the Coulomb phase are either
one (three vacua with unbroken gauge group $\uuu\times\u$), two (the
two $\uu\times\uu$ vacua) or three (three $\u\times\uuu$ vacua). This 
structure is manifest on the quantum equation satisfied by $u$
\cite{fer4},
\be\label{Qeq} (u-1)^{3}(u-2)^{2}(u-3)^{3} + \frac{\la^{4}}{64} = 0\,
.\ee
Varying $\la$, it is straightforward to show that the eight roots of
(\ref{Qeq}) can be permuted, and thus that the eight Coulomb sheets
are glued together by branch cuts. The full $\mathcal M_{\mathrm q}$
can be found by repeating this kind of argument and studying the
possible phase transitions \cite{fer4}. In \cite{CSW}, it was
emphasized that the smooth interpolations can connect vacua with
different patterns of gauge symmetry breaking. In our framework, this 
is a direct consequence of (\ref{Qeq}) for example.
\vfill\eject
\subsection{Conclusions}

Our main result was to show that D-branes can be smoothly deformed
into solitonic branes and vice versa. We have also discovered new
extended objects, with tensions scaling as fractional powers of the
string coupling.

D-branes are simple in string theory but are hard to study in gauge
theory. On the other hand, we have explained that solitonic branes can
be quantized semiclassically in gauge theory. This provides a
microscopic description of solitonic branes using gauge theories,
which is the counterpart of the microscopic description of D-branes
using open strings. Since a microscopic description of solitonic
branes in string theory has remained elusive, this approach is
probably worth pursuing.

\section{Planar diagrams and Geometry}
\subsection{Calabi-Yau geometries}

A famous non-compact Calabi-Yau geometry is the conifold $\mathcal
O(-1)\oplus\mathcal O(-1) \rightarrow \pone$. By wrapping $N$ D5
branes on the $\pone$, we engineer the pure $\nn=1$, $\uN$ gauge
theory \cite{MN} with a Yang-Mills coupling
\be\label{gYM} g^{2}_{\mathrm{YM}}\sim
\frac{1}{\mathrm{vol}(\pone)}\,\cdotp\ee
To add adjoint chiral multiplets, we need to consider a
more general local geometry
\be\label{lgeom}\mathcal O(-1+M)\oplus\mathcal O(-1-M) 
\rightarrow \pone\, .\ee
This space is Calabi-Yau because the first Chern class $2$ of the
$\pone$ and $-1-M-1+M=-2$ of the normal bundle compensate each other.
It is not difficult to check, by direct analysis or by using the
Riemann-Roch theorem, that the geometry (\ref{lgeom}) has a
$M$-parameter family of holomorphic two-spheres. These $M$ parameters 
yield $M$ adjoint chiral multiplets $X_{i}$ in the world-volume
gauge theory of D-branes wrapped on one of these two-spheres.

The existence of an $M$ dimensional continuous family of $\pone$s
implies that there is no superpotential for the $X_{i}$. To turn on a
$W$, we generalize the geometry (\ref{lgeom}) to a new Calabi-Yau
$\mathscr M$, covered by two coordinate patches $(z,w_{1},w_{2})$ and
$(z',w_{1}',w_{2}')$, and defined by the transition functions
\cite{fer2}
\be\label{tf}\mathscr M:\, \left\{
\begin{matrix}
\,\, z'=1/z\hfill\ \\ w_{1}' = z^{1-M}w_{1}\hfill \ \\
w_{2}' = z^{1+M}w_{2} + \partial_{w} E (z,w_{1})\, .\hfill \ 
\end{matrix}\right.\ee
When $E=0$, we find (\ref{lgeom}), but in general the perturbation $E$
is a non-zero function of two complex variables. It can be Laurent
expanded in terms of entire functions (often polynomials)
$E_{j}$ in the form
\be\label{Edef} E(z,w) = \sum_{j=-\infty}^{+\infty}E_{j}(w)z^{j}\,
.\ee

From general arguments \cite{math}, it is known that perturbing the
geometry (\ref{lgeom}) induces an obstruction to the space of versal
deformations of the $\pone$s. Holomorphic $\pone$s can only exist at
discrete values of the $X_{i}$ that are solutions of $M$ constraints
(interestingly, the fact that the number of constraints is equal to
the number of parameters is equivalent to the Calabi-Yau condition).
In the case of (\ref{tf}), the constraints can be
put in the form $\d W=0$ for a certain holomorphic function
$W(X_{1},\ldots,X_{M})$. Turning on $E$ is the most general known
perturbation having this property. Obviously, $W$ is the tree-level
superpotential of the world volume gauge theory.

\subsection{Superpotential}

A formula to compute the superpotential in a variety of cases where
branes are wrapped on two-cycles $\Sigma$ was given 
by Witten in \cite{Witwalls},
\be\label{Wsup1} W(\Sigma) = \int_{B}\Omega + \mathrm{constant}\, .\ee
The chain $B$ is such that $\partial B=\Sigma - \Sigma_{0}$ for some
fixed two-cycle $\Sigma_{0}$, and $\Omega$ is the holomorphic
three-form. It is not difficult to check that the equations $\d W=0$
yield the correct conditions for $\Sigma$ to be holomorphic. In our
case, $\Sigma$ is a two-sphere in $\mathscr M$, parametrized by the
$X_{i}$, and computing the integral (\ref{Wsup1}) yields
$W(X_{1},\ldots,X_{M})$. This is perfectly good as long as we consider
a single wrapped brane. When dealing with two or more branes, the
$X_{i}$ are non-commuting matrices and (\ref{Wsup1}) becomes
ambiguous. The correct procedure is then to write directly the
conditions for supersymmetry to be preserved. The $M$ equations for
non-commuting variables so obtained can still be derived by
extremizing a superpotential $W$ \cite{fer2}
\be\label{Wfor} W=
\frac{1}{2i\pi}\oint_{C_{0}}\! z^{-M-1}\tr E\Bigl(z,
\sum_{j=1}^{M}X_{j}z^{j-1}\Bigr) \d z\, .\ee
The contour integral is over a small loop $C_{0}$ encircling
the origin. In the commutative limit, (\ref{Wfor}) and (\ref{Wsup1})
are of course equivalent.

For example, in the case $M=1$, (\ref{Wfor}) yields $W(X) = \tr
E_{1}(X)$. In the case $M=2$, we can consider an arbitrary
superpotential $W(x,y) = \sum_{i,j\geq 0}a_{ij}x^{i}y^{j}$ in the
commutative limit. The formula (\ref{Wfor}) gives the ordering
prescription for any given monomial,
\be\label{Weyl}x^{i}y^{j}\mapsto\frac{i!j!}{(i+j)!}\oint_{C_{0}}\frac{\d
z}{2i\pi} z^{-1-j} (X+Y z)^{i+j}\, .\ee
This is the standard Weyl ordering.

The superpotential (\ref{Wfor}) includes as special cases all the
previously engineered superpotentials for gauge theory with adjoints
(see for example \cite{CKV}). The associated multi-matrix integrals
(\ref{MM}) are highly non-trivial and far beyond the grasp of ordinary
matrix model techniques. Fortunately, we now have an entirely new
point of view on the problem: the solution of the model amounts to
finding the closed string background $\hat{\mathscr M}$ dual to the
brane theory on $\mathscr M$. We shall now explore this strategy, by
describing the two main steps required in the computation of the space
$\hat{\mathscr M}$.
\eject
\subsection{Blowing down}

Holomorphic two-spheres in the geometry (\ref{tf}) are associated with
the critical points of the superpotential (\ref{Wfor}). For a generic
Morse critical point, the adjoint fields are massive and can be
integrated out. This means that the normal bundle to the associated
$\pone$ is $\mathcal O(-1)\oplus\mathcal O(-1)$ as in the case of the
pure $\nn=1$ theory. More generally, if the corank of the Hessian of
$W$ at the critical point is $r$, we expect the normal bundle to be
$\mathcal O(-1+r)\oplus\mathcal O(-1-r)$, corresponding to a massless
theory with $r$ adjoint multiplets. This was checked explicitly in the
case $M=2$ in \cite{fer2}. When $r\leq 2$, the low energy theory is
asymptotically free, and we may expect that quantum mechanically the
$\pone$ shrinks to zero size. This heuristic RG argument is in
beautiful agreement with a theorem by Laufer \cite{Lau} that states
that shrinkable spheres with normal bundle $\mathcal
O(-1+r)\oplus\mathcal O(-1-r)$ must have $r=0,1$ or 2.

The blow down map 
\be\label{bdm}\pi :\ \mathscr M\rightarrow{\mathscr M}_{0}\ee
is a birational isomorphism except on the $\pone\subset\mathscr M$
that are mapped onto singular points of ${\mathscr M}_{0}$. Blowing
down is a mathematically precise way to implement the heuristic
picture of shrinking spheres. The calculation of the blow down map,
for given $\mathscr M$, is a non-trivial problem. In the simplest
cases, it amounts to finding four globally holomorphic functions
$\pi_{i}(z,w_{1},w_{2})\in\mathbb C$ on $\mathscr M$ that maps the
$\pone$ onto points. Global holomorphicity is checked by using the
transition functions (\ref{tf}). The four functions $\pi_{i}$ depend
on three variables and satisfy in general an algebraic equation that
defines the singular variety ${\mathscr M}_{0}$ as a hypersurface in
$\mathbb C^{4}$. Several explicit examples can be found in
\cite{fer2}.

In the geometric transition picture, the singular space ${\mathscr
M}_{0}$ describes the classical $S\rightarrow 0$ matrix model
(\ref{MM}), or equivalently the classical equations of motion
\be\label{cleq}\d W = \tr\d V(X_{i}) = 0\, .\ee
When the matrices $X_{i}$ are commuting (one dimensional
representations), (\ref{cleq}) reduces to a set of algebraic
equations. By construction of the blow down map, these solutions are
automatically associated with singular points in ${\mathscr M}_{0}$.
More generally, the correspondence with the matrix model implies that the
higher dimensional representations of (\ref{cleq}) must also be
associated with singular points on ${\mathscr M}_{0}$. This means that
classical algebraic geometry must know about the non-commutative
structure that emerges when considering D-branes.

Let us illustrate this point on a non-trivial example \cite{fer2}. We 
consider the geometry
\be\label{tfex}\mathscr M:\, \left\{
\begin{matrix}
\,\, z'=1/z\hfill\ \\ w_{1}' = z^{-1}w_{1}\hfill \ \\
w_{2}' = z^{3}w_{2} + w_{1}^{2} - z G(w_{1}^{2}/z^{2}) \\&\hskip -3cm
- z^{2}F_{1}(w_{1}^{2}) - z^{2}w_{1}F_{2}(w_{1}^{2})\, .\hfill \ 
\end{matrix}\right.\ee
The corresponding superpotential is
\be\label{Wex} W(X,Y) = \tr\bigl( XY^{2} + A(X) + B(Y) \bigr)\, ,\ee
with $A'(X) = -F_{1}(X^{2}) - XF_{2}(X^{2})$ and $B'(Y) = -G(Y^{2})$.
The classical equations of motion are
\be\label{cleqex} \{ X,Y\} = G(Y^{2})\, ,\ Y^{2} = F_{1}(X^{2}) + 
XF_{2}(X^{2})\, .\ee
These equations can be analysed by noting that $X^{2}$ and $Y^{2}$ are
Casimir operators. We find that there are $\max (\deg A' + 2,2\deg A'
\deg G)$ one dimensional irreducible representations and $[\frac{1}{2}
(\deg A' -1)]$ two dimensional irreducible representations (for which
$X$ and $Y$ are linear combinations of Pauli matrices). On the other
hand, the blow down map can be explicitly constructed \cite{fer2} and
the singular Calabi-Yau is found to be
\begin{multline}\label{M0ex}\hskip -.25cm{\mathscr M}_{0}:\, x_{2}\bigl[
(x_{2}-F_{1}(x_{1}))^{2} - x_{1}F_{2}^{2}(x_{1})\bigr] = x_{4}^{2} 
- x_{1}x_{3}^{2}\\ - G(x_{2})\bigl[ (x_{2}-F_{1}(x_{1}))x_{3} 
+ x_{4}F_{2}(x_{1})\bigr]\, .\end{multline}
It is not difficult to check that ${\mathscr M}_{0}$ has
singular points for both one {\it and} two dimensional representations.

The fact that the full set of solutions of matrix equations can be
encoded in a set of algebraic equations describing the singularities
of a Calabi-Yau is a remarkable feature. In all the cases that have
been worked out, the solutions to the matrix equations can be
characterized by a set of Casimir operators. These Casimirs turn out
to be associated with some coordinates on ${\mathscr M_{0}}$. The
special values of the Casimirs labeling the solutions then match the
values of the coordinates at the singular points. For example, in the 
case (\ref{M0ex}), the coordinate $x_{1}$ is associated with the
Casimir $X^{2}$ and the coordinate $x_{2}$ with the Casimir $Y^{2}$.

\subsection{Deforming}

The full $S\not = 0$ matrix model (\ref{MM}) is described by a
deformed non-singular geometry $\hat{\mathscr M}$ whose equation is
obtained from the equation for ${\mathscr M}_{0}$ by adding certain
monomials $\sum c_{abcd}x_{1}^{a}x_{2}^{b}x_{3}^{c}x_{4}^{d}$. The
allowed monomials must be such that the divergences in the period
integrals $\oint\Omega$ of the holomorphic three-form over non-compact
cycles are either $S$-independent of linear in $S$ and logarithmic.
This condition is the mathematical counterpart of the renormalization
properties of the $\nn=1$ Yang-Mills theories. After implementing this
constraint, and taking into account possible coordinate redefinitions,
we obtain a certain number ${\mathscr N}_{\mathrm{geo}}$ of
independent and freely adjustable coefficients $c_{abcd}$ that
parametrize the geometry $\hat{\mathscr M}$.

The parameters $c_{abcd}$ should not be confused with the parameters
in the superpotential or equivalently in the original geometry
(\ref{tf}). They have a very natural interpretation in terms of the
matrix model. The matrix integrals (\ref{MM}) are defined by summing
over planar Feynman diagrams. The Feynman rules are themselves defined
by expanding around a particular solution of the classical equations
of motion. The most general classical solution is the direct sum of
the irreducible representations of (\ref{cleq}), and is thus
characterized by parameters, called filling fractions, giving the
fraction of eigenvalues in given irreducible representations. The
number ${\mathscr N}_{\mathrm{alg}}$ of filling fractions that need to
be specified is thus equal to the number of irreducible
representations of (\ref{cleq}). Consistency requires that there
should exist a one-to-one mapping between the filling fractions and the
$c_{abcd}$, and thus that
\be\label{geomalg}{\mathscr N}_{\mathrm{geo}} = {\mathscr
N}_{\mathrm{alg}}\, .\ee
This is a profound equality relating classical geometry and
non-commutative algebra: the left-hand side is computed by studying
the ``renormalizable'' deformation of a certain singular Calabi-Yau
space, and the right-hand side is computed by working out the
representation theory of a certain matrix algebra. This correspondence
follows rather straightforwardly from the physical picture, but
remains to be understood at a deeper mathematical level. It was
checked explicitly in a variety of non-trivial examples in
\cite{fer2}. For example, in the theory
(\ref{tfex},\ref{Wex},\ref{M0ex}), detailed calculations show that
\begin{multline}\label{NNex}
{\mathscr N}_{\mathrm{alg}} = \max\bigl(\deg A'+2,2\deg
A'\deg G\bigr) \\ +\bigl[(\deg A' -1)/2\bigr] =
{\mathscr N}_{\mathrm {geo}}\end{multline}
as expected.

\subsection{The solution}

In the case of the one-matrix model $W=\tr V(X)$, it is well-known
that the deformed geometry takes the simple form
\be\label{Mh1}\hat{\mathscr M}:\
x_{4}^{2}=x_{3}^{2}+x_{2}^{2}-V'(x_{1})^{2}+S\Delta (x_{1})\, ,\ee
where $\Delta$ is a polynomial of degree $\deg V'-1$ describing the
deformation of ${\mathscr M}_{0}$ (in this case it is obvious that
$\mathscr N_{\mathrm{alg}} = \deg V'$; it is also straightforward to
show that $\mathscr N_{\mathrm{geo}} = \deg V'$ by computing the
periods $\oint\Omega$). It is useful to think about (\ref{Mh1}) as a
fibration over a base parametrized by $x_{1}$. The fiber $F_{x_{1}}$
at $x_{1}$ is the simplest ALE space $\mathbb C^{2}/\mathbb Z_{2}$,
and contains a single holomorphic two-sphere $S^{2}$. The
discontinuity of the resolvent
\be\label{gdef} g^{X}(x) =
S\Bigl\langle\frac{\tr}{n}\frac{1}{x-X}\Bigr\rangle\ee
across its branch cuts is then simply given by
\be\label{disc}g^{X}(x+i\epsilon)-g^{X}(x-i\epsilon) =
\int_{S^{2}\subset F_{x}}\mathrm i_{X}\Omega\, ,\ee
where $\mathrm i_{X}$ denotes the interior product with respect to the
vector field $\partial/\partial x_{1}$. 

More general multi-matrix models can always be reduced to one matrix
models with a complicated effective potential after integration over
all but one matrix. This shows that the structure just described is
still relevant, but the fibers $F_{x}$ are typically much more
complicated than $\mathbb C^{2}/\mathbb Z_{2}$, and contains many
two-spheres (as a consequence of the multivaluedness of the effective
potential). The basic formula (\ref{disc}) is still valid \cite{fer2}
and can be used to find explicitly the resolvents of the various
matrices of multi-matrix models. For example, in the case of
(\ref{Wex}), it is shown with this method that $g^{X}$ satisfies a
degree $\max(3,2+2\deg G)$ algebraic equation and that $g^{Y}$ lies at
the intersection of $d$ quadrics in a $d$-dimensional auxiliary space
\cite{fer2}. It appears to be very difficult to obtain these results
using standard matrix model technology. Nevertheless, in the special
case where $G$ is a constant, the loop equations can be solved
\cite{fer2} and full agreement is found with the geometry.

\subsection{Conclusions}

We have a rich interplay between algebraic geometry and matrix
models, that suggests many non-trivial results. There are many open
problems: can we understand the special properties of the matrix
models with potentials (\ref{Wfor}) (irreducible representations of
the classical equations of motion, loop equations, etc...)?; does the
potential (\ref{Wfor}) describes the most general matrix model that
can be engineered in string theory?; can we give a direct mathematical
proof of the relation between the singularity structure of ${\mathscr 
M}_{0}$ and the representations of (\ref{cleq})?; and so on. 
The most important
question is probably to understand whether the blow down map for the
geometry (\ref{tf}) always exists. This is a hard mathematical problem
\cite{kol}. If it does, and if we can find an algorithm to
compute it, then we have solved a huge class of multi-matrix models
(\ref{Wfor}). If it does not, then the picture of geometric
transitions cannot deal with our examples in general, with fundamental
consequences for our understanding of the structure of $\nn=1$ vacua
in string theory.

\section{Homework}
\subsection{Exercices}

\noindent 
a) Prove or disprove that the $\nn=1$ quantum parameter spaces are always
connected.

\noindent
b) Describe the correct microscopic description of solitonic branes
using gauge theory.

\subsection{Problem}

Consider the geometry (\ref{tf}).

\noindent i) Find the blow down geometry. Compare with the algebra $\d
W=0$ where $W$ is given by (\ref{Wfor}).

\noindent ii) Find the deformed geometry. Check (\ref{geomalg}).

\noindent iii) Compute the resolvents from the geometry. Compare with 
the loop equations.





\section*{Acknowledgements} I wish to thank Costas Bachas, Eug\`ene
Cremmer and Paul Windey for organizing a wonderful Strings 2004
conference at Coll\`ege de France. My work was supported in part by
the Swiss National Science Foundation, by the belgian Institut
Interuniversitaire des Sciences Nucl\'eaires (convention 4.4505.86),
by the Interuniversity Attraction Poles Programme (Belgian Science
Policy) and by the European Commission RTN programme HPRN-CT-00131 (in
association with K.\ U.\ Leuven). I am on leave of absence
from Centre National de la Recherche Scientifique, Laboratoire de
Physique Th\'eorique de l'\'Ecole Normale Sup\'erieure, Paris, France.

\end{twocolumn}
\end{document}